\documentclass{aastex63}
\usepackage{multirow}
\usepackage{booktabs}

\received{xxx}
\revised{yyy}
\accepted{zzz}

\submitjournal{ApJ}

\begin{document}

\title{ The \lq{canonical\rq}  White Dwarf Cooling Sequence of M5
}

\author[0000-0002-8004-549X]{Jianxing Chen}
\affiliation{Dipartimento di Fisica e Astronomia ``Augusto Righi'',
  Alma Mater Studiorum Universit\`a di Bologna, via Piero Gobetti
  93/2, I-40129 Bologna, Italy}
\affiliation{INAF-Osservatorio di Astrofisica e Scienze dello Spazio
  di Bologna, Via Piero Gobetti 93/3 I-40129 Bologna, Italy}
%
%
\author[0000-0002-2165-8528]{Francesco R. Ferraro}
\affiliation{Dipartimento di Fisica e Astronomia ``Augusto Righi'',
  Alma Mater Studiorum Universit\`a di Bologna, via Piero Gobetti
  93/2, I-40129 Bologna, Italy}
\affiliation{INAF-Osservatorio di Astrofisica e Scienze dello Spazio
  di Bologna, Via Piero Gobetti 93/3 I-40129 Bologna, Italy}
\author[0000-0002-2744-1928]{Maurizio Salaris}
\affiliation{Astrophysics Research Institute, Liverpool John Moores
  University, Liverpool Science Park, IC2 Building, 146 Brownlow Hill,
  Liverpool L3 5RF, UK}

\author[0000-0002-5038-3914]{Mario Cadelano}
\affiliation{Dipartimento di Fisica e Astronomia ``Augusto Righi'',
  Alma Mater Studiorum Universit\`a di Bologna, via Piero Gobetti
  93/2, I-40129 Bologna, Italy}
\affiliation{INAF-Osservatorio di Astrofisica e Scienze dello Spazio
  di Bologna, Via Piero Gobetti 93/3 I-40129 Bologna, Italy}
\author[0000-0001-5613-4938]{Barbara Lanzoni}
\affiliation{Dipartimento di Fisica e Astronomia ``Augusto Righi'',
  Alma Mater Studiorum Universit\`a di Bologna, via Piero Gobetti
  93/2, I-40129 Bologna, Italy}
\affiliation{INAF-Osservatorio di Astrofisica e Scienze dello Spazio
  di Bologna, Via Piero Gobetti 93/3 I-40129 Bologna, Italy}

\author[0000-0002-7104-2107]{Cristina Pallanca}
\affiliation{Dipartimento di Fisica e Astronomia ``Augusto Righi'',
  Alma Mater Studiorum Universit\`a di Bologna, via Piero Gobetti
  93/2, I-40129 Bologna, Italy}
\affiliation{INAF-Osservatorio di Astrofisica e Scienze dello Spazio
  di Bologna, Via Piero Gobetti 93/3 I-40129 Bologna, Italy}

\author[0000-0002-7104-2107]{Leandro G. Althaus}
\affiliation{Grupo de Evolucion Estelar y Pulsaciones, Facultad de Ciencias Astronómicas y Geofisicas, Universidad Nacional de La Plata,
Paseo del Bosque s/n, 1900 La Plata, Argentina}
\affiliation{CCT – CONICET Centro Científico Tecnológico La Plata, Consejo Nacional de Investigaciones Científicas y Técnicas,
Calle 8 No. 1467, B1904CMC La Plata, Buenos Aires, Argentina}

\author[0000-0001-5870-3735]{Santi Cassisi}
\affiliation{INAF-Osservatorio Astronomico d'Abruzzo, Via Maggini, I-64100 Teramo, Italy}
\affiliation{INFN - Sezione di Pisa, Largo Pontecorvo 3,   I-56127 Pisa, Italy}
%

\begin{abstract}
Recently, a new class of white dwarfs (dubbed ``slowly cooling WDs'')
has been identified in two globular clusters (namely M13 and NGC 6752)
showing a horizontal branch (HB) morphology with an extended blue
tail.  The cooling rate of these WDs is reduced by stable
thermonuclear hydrogen burning in their residual envelope, and they
are thought to be originated by stars that populate the blue tail of
the HB and then skip the asymptotic giant branch phase.  Consistently,
no evidence of such kind of WDs has been found in M3, a similar
cluster with no blue extension of the HB.  To further explore this
phenomenon, we took advantage of deep photometric data acquired with
the Hubble Space Telescope in the near-ultraviolet and investigate the
bright portion of the WD cooling sequence in M5, another Galactic
globular cluster with HB morphology similar to M3.  The normalized WD
luminosity function derived in M5 turns out to be impressively similar
to that observed in M3, in agreement with the fact that the stellar
mass distribution along the HB of these two systems is almost
identical. The comparison with theoretical predictions is consistent
with the fact that the cooling sequence in this cluster is populated
by canonical (fast cooling) WDs.  Thus, the results presented in this
paper provide further support to the scenario proposing a direct
causal connection between the slow cooling WD phenomenon and the
horizontal branch morphology of the host stellar cluster.
\end{abstract}


\section{Introduction}
\label{sec:intro}
White dwarfs (WDs) constitute the final evolutionary stage of stars
with low and intermediate initial mass (below $8 M_\odot$, possibly up
to $11 M_\odot$; e.g., \citealp{Corsico+2019}). This mass range indeed
includes the overwhelming majority ($\sim 98\%$; e.g.,
\citealt{Winget+2008, Woosley+2015}) of stars in our universe.  As the 
ending point of stellar evolution, WDs gather tremendous information
about the evolutionary history of their progenitors.  In addition,
they are characterized by relatively simple structures, thus providing
the ideal laboratory for studying the behaviour of matter in extreme
high-density physical conditions \citep{Garcia-Berro+1995}.
 
WDs are thought to be the remnant core of stars that have terminated
their thermonuclear activity and have released their outer layers. Because
of the lack of energy production
the evolution of WDs is generally described
as a cooling process. Thus, WD aging is essentially
characterized by a progressive cooling and fading at roughly fixed radius as a
function of time (the so-called \lq{cooling time\rq}). This provides a
natural link between WD luminosities and their cooling ages, which is
often adopted as a cosmic chronometer to measure the age of stellar
populations in the Galaxy (as the Galactic disk, globular and open
clusters; see e.g., \citealt{Hansen+2007, Bellini+2010, Bedin+2010,
  Bedin+2023, Kilic+2017}).
 
However, recent theoretical models \citep{Althaus+2015} have shown
that small extra amounts of residual hydrogen (a few $10^{-4} M_\odot$)
left on the WD surface from the previous evolutionary stages is
sufficient to allow quiescent thermonuclear burning. This process
provides a non-negligible source of energy that can significantly
decrease the cooling rate with respect to that of normal WD,
especially in the regime of low-mass (about $<0.56 M_\odot$) WDs and low-metallicity
($Z<0.001$) progenitors \citep{Renedo+2010, Miller+2013}, thus generating a
new class of WDs that we named \lq{slowly cooling WDs\rq}.  The cooling
time-scale increase is expected to enhance the number of WDs in the
brightest and intermediate portions of the cooling sequence, with an observable
modification of the shape of the WD luminosity function (LF).
In turn, this implies 
that the effect of possible residual hydrogen burning cannot be
neglected when \lq{reading\rq} the cooling time (hence, the WD age) along
the cooling sequences, expecially in Galactic globular clusters (GGCs)
that are the oldest and most metal-poor stellar populations in the
Milky Way (thus hosting populations of low-mass WDs
produced by low-metallicity progenitors).

The first observational evidence of the existence of slowly cooling
WDs
has been presented by \citet{Chen+2021} who compared the WD LFs of the 
\lq{twin clusters\rq} M13 and M3, discovering a significant WD excess
in the former.  These two clusters share many physical properties
(e.g., the metallicity, age, central density, etc.;
\citealp{Harris+1996, Dotter+2010}), with the notable exception of the
horizontal branch (HB) morphology, which is completely different in
the two systems \citep{Ferraro+1997a, Dalessandro+2013a}. In fact, the
HB morphology in M13 is characterized by an extended blue tail (see
also \citealt{Ferraro+1998}), while no blue extension is present in
the HB of M3 \citep{Buonanno+1994, Ferraro+1997b}. \citet{Chen+2021}
confirmed that the HB morphologies are related to the presence or lack
thereof of slowly cooling WDs, because they correspond to quite
different mass distributions: stellar masses increase from the blue
edge to the red edge of the HB, implying that most of the HB stars in
M13 are less massive than that in M3.  On the other hand, the stellar
mass in the HB phase is the key parameter that sets the subsequent
evolution. In fact, theoretical HB
models demonstrate that extremely low-mass HB stars
(less massive than $0.56 M_\odot$ at
intermediate metallicities) skip 
the thermal pulses and associated third dredge-up, a
mixing process occurring during the Asymptotic Giant Branch (AGB),
which is the evolutionary phase immediately following the HB and
preceding the WD stage. This process efficiently mixes the material
present in the envelope of AGB stars with hotter inner layers, bringing hydrogen into
deeper regions where it is burned. As a consequence, a star
experiencing the third dredge-up reaches the final WD stage  
with almost no residual hydrogen envelope, and no efficient hydrogen burning.
Conversely, HB stars less massive than $0.56 M_\odot$
skipping this critical event, reach the WD stage with a residual
hydrogen envelope thick enough to allow stable thermonuclear burning,
which provides the WD with an extra-energy production that delays
its cooling \citep{Althaus+2015}.
Thus, while all the (red) HB stars in M3 are more massive
than this value and are therefore expected to experience the third
dredge-up in the AGB phase, the majority of the stars in M13 populating
the HB blue tail (hence less massive than the mass threshold) are predicted to skip this event and thus produce
slowly cooling WDs. This scenario has been
fully confirmed by \citet{Chen+2022} in the case of NGC 6752, 
a GGC with almost the same metallicity and the same (extended blue) HB
morphology of M13.
The results obtained in M13 and NGC 6752 therefore provided firm
evidence for the existence of slowly cooling WDs and solid empirical
support to the predicted physical origin of these objects (i.e., the
fact that their progenitors skip the third dredge-up). This also
consolidated the connection between the occurrence of this phenomenon
and the HB morphology: significant populations of slowly cooling WDs
are expected in GCs with a HB displaying a
well populated and extended blue-tail.

To further verify the scenario in the case of a cluster with no HB
blue extension, in this paper we use deep near-ultraviolet photometry
secured with the HST to investigate the LF of the brightest portion of
the WD cooling sequence in M5. This is a luminous ($M_V=-8.8$) and
well studied GGC \citep[e.g.][]{Lanzoni+2007, Lanzoni+2016,
  Lanzoni+2018, Ferraro+2012, Miocchi+2013, Pallanca+2014,
  Piotto+2015, Nardiello+2018}, relatively close to the Sun (at a
distance of $\sim 7.5$ kpc). The HB morphology of M5 is similar to
that of M3, with no blue tail extension, characterized by a rich
population of RR Lyrae \citep{Arellano+2016a,
  Arellano+2016b}. According to the scenario discussed above this
should imply that most of the HB stars in M5 experience the third
dredge-up and end their evolution as canonical WDs.
 
%
%
The paper is organized as follows: in Section \ref{sec:data}, we
describe the data reduction and artificial star tests; the WD sample
selection and analysis are presented in Section \ref{sec:analysis};
the discussion and conclusions of the work are provided in the Section
\ref{sec:con}.


\section{Data reduction}
\label{sec:data}

In this work, we use the deep and high-resolution photometric data in
the near-ultraviolet band, obtained with the HST/WFC3 camera in the
UVIS channel. The dataset has been acquired as a part of the HST Large
Legacy Treasury Program (GO-13297, PI: Piotto; see
\citealp{Piotto+2015} and it is composed of 4 frames (two with
exposure time $t_{\rm exp}=689$s, two with $t_{\rm exp}= 690$) in the
F275W filter, and 4 frames (each with $t_{\rm exp}=306$s) in the F336W
filter.

For the photometric analysis we used UVIS exposures with $``\_flc"$
extension, which are calibrated and corrected for Charge Transfer
Efficiency (CTE). After the pre-reduction procedure, which includes
the extraction of the science images (chip1 and chip2) from the raw
fits files and the application of the Pixel Area Map correction, in
order to derive the star magnitudes we performed a Point Spread
Function (PSF) fitting following the so-called ``UV-route''.  This is
a photometric procedure specifically optimized for the detection of
blue and hot objects, such as hot HB stars, blue straggler stars, and
WDs in crowded fields \citep[see, e.g.,][]{Ferraro+1997a,
  Ferraro+1997b, Ferraro+2001, Ferraro+2003, Ferraro+2018,
  Lanzoni+2007, Dalessandro+2013a, Dalessandro+2013b, Raso+2017}).  In particular, \citet{Raso+2017} discussed the net advantages of using
this {\it UV-Guided} search of the sources with respect to the
standard {\it optical-driven} selection, in order to derive complete
samples of blue and faint objects in high-density old clusters, where
the optical emission is primarily dominated by a large population of
bright and cool ($3500-5000$ K) giants.  Specifically, in this paper
we followed the approach described in \citet{Cadelano+2019} and
\citet[][but see also \citealt{Cadelano+2017a, Cadelano+2022a, Raso+2017, Raso+2020, Chen+2021,
    Chen+2022}]{Cadelano+2020a}. The main idea of the procedure is to
created a master list of sources detected in the near-ultraviolet
images and then force the detection in images acquired at longer
wavelengths (as blue and optical bands).

In short, we first selected about $\sim 250$ bright and unsaturated
stars uniformly distributed in the entire field of view to properly
model the PSF function for each exposure. We then applied the
resulting model to all the sources detected above $5\sigma$ from the
background, and combined the stars appearing in at least 2 images,
thus building the master list. Secondly, we performed the \lq{forced\rq} 
photometry with DAOPHOT/ALLFRAME \citep{Stetson+1994} on all the
images at positions corresponding to the stars in the master list,
even in a location where the signal is below $5\sigma$. Finally, the
magnitudes of all stars in each filter were homogenized, and the
magnitudes and corresponding errors were calculated from their
weighted mean and standard deviation.

After correcting the instrumental positions for geometric distortions,
we calibrated instrumental magnitudes to the VEGMAG system and aligned
the instrumentl coordinates to the International Celestial Reference
System by cross-correlatioind with the reference catalog of the HST UV
Globular Cluster Survey \citep{Piotto+2015, Nardiello+2018}. We also
executed an additional visual inspection to clean up the sample from
fake sources caused by saturation of extremely bright stars.  As a
result of the data reduction process, a final catalog listing almost
40,000 stars has been finally obtained.

\begin{figure}
\centering
	\includegraphics[width = 12.6cm]{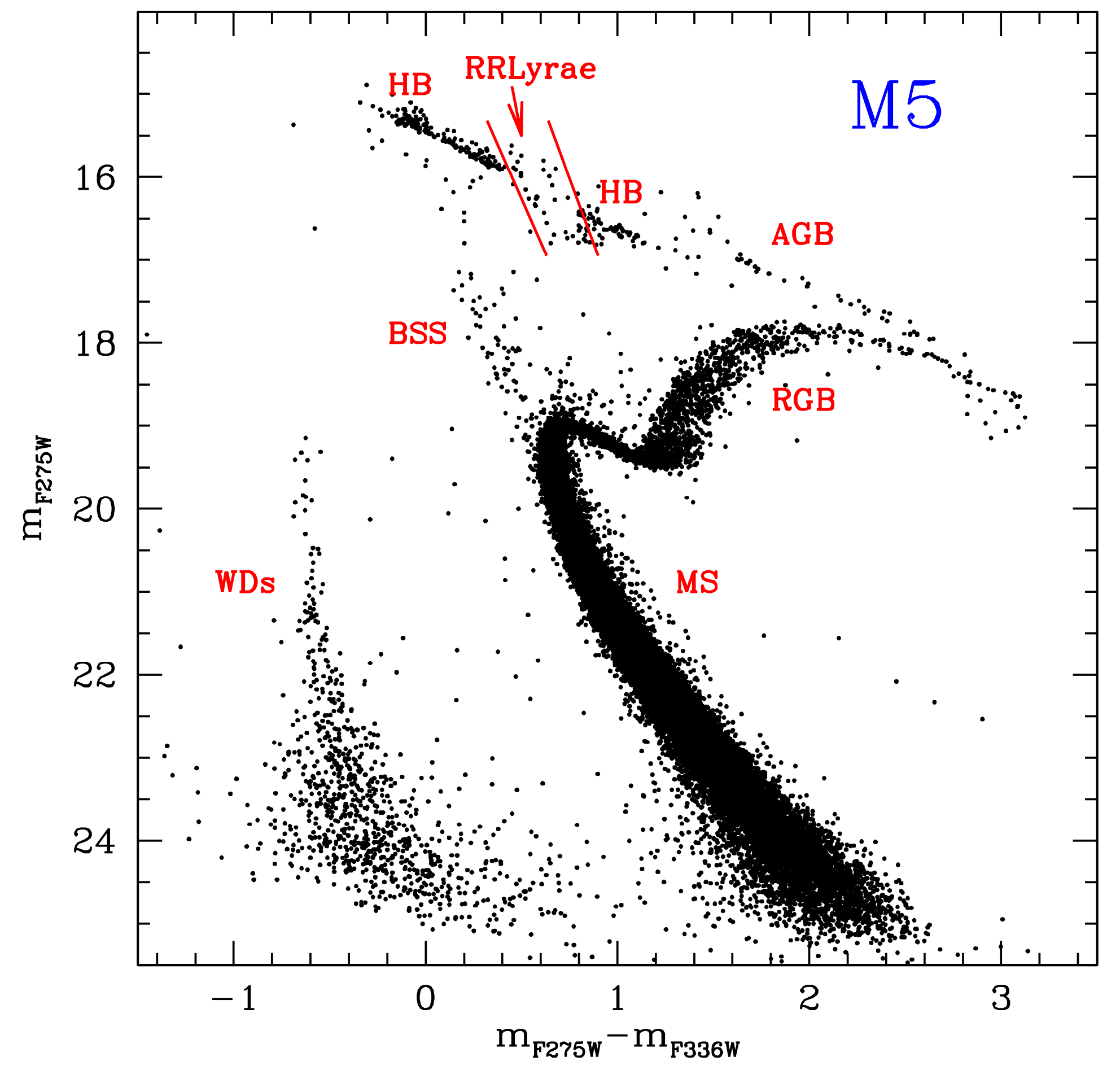}
	\caption{The near-ultraviolet CMD of the 39114 stars detected in M5. The various stellar populations are indicated.
The two red lines delimitate the CMD region where most of the RR Lyrae are found. }
	\label{fig:1}
\end{figure}
\subsection{Color-magnitude diagram}
The color-magnitude diagram (CMD) obtained for M5 is presented in
Figure \ref{fig:1}. All the main evolutionary
sequences are clearly defined and they appear negligibly affected by 
contamination from field stars or background extra-Galactic
sources. The CMD extends from $m_{\rm F275W}=15$ to $m_{\rm
  F275W}\simeq 25$, spanning more than 10 magnitudes, and 
provides a panoramic over-view of the stellar population hosted in
the cluster. As expected, the near-ultraviolet CMD appears
significantly different from the \lq{classical\rq} optical one:
the luminosity of red giant branch (RGB) stars is significantly
suppressed (the RGB-tip is at $m_{\rm F275W}\sim 19$), and the same
holds for the AGB, which is clearly distinguishable from the RGB in
the color range $1.6<(m_{\rm F275W}-m_{\rm F336W})<3$. The luminosity
in the F275W band is instead dominated by HB stars, clearly visible at
$(m_{\rm F275W}-m_{\rm F336W})<1.2$.  A well defined sequence of blue
straggler stars (see also \citealp{Lanzoni+2007}) is seen emerging
from the main sequence turn-off (MS-TO, at $m_{\rm F275W}\sim 19$) and
extends up to the HB at $m_{\rm F275W}\sim 16$.  A copious population
of RR Lyrae variables (observed at random phases) is also visible as a
sort of stream of stars diagonally crossing the HB at $0.5<(m_{\rm
  F275W}-m_{\rm F336W})<1$.  The brightest portion of the WD cooling
sequence appears as a well-defined and populated, almost vertical
sequence at $(m_{\rm F275W}-m_{\rm F336W})\sim-0.8$, extending for
almost 6 magnitudes from $m_{\rm F275W}\sim 25.5$ up to $m_{\rm F275W}\sim 19$.

\begin{figure}
\centering
	\includegraphics[width = \textwidth]{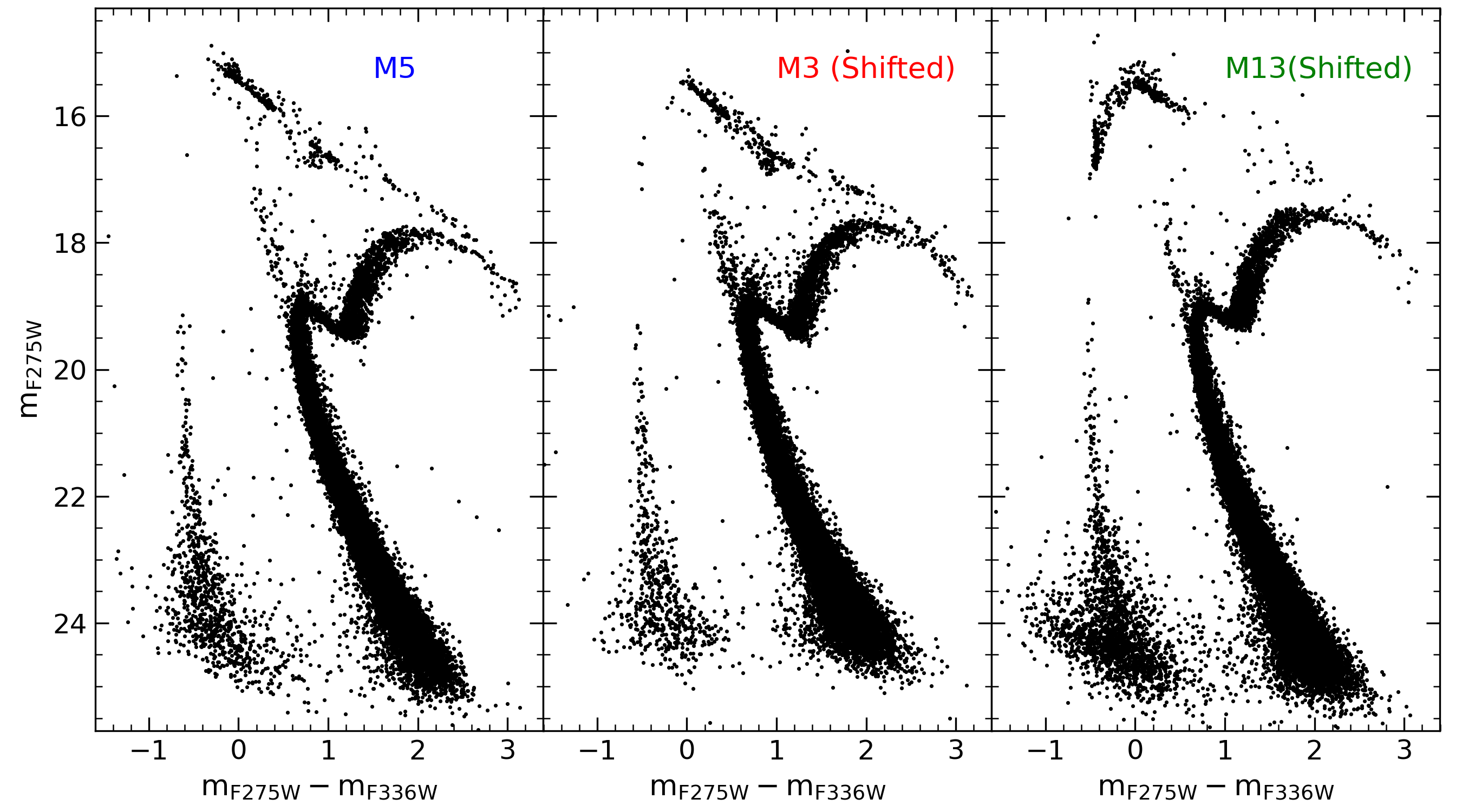}
	\caption{CMD of M5 {\it (left panel)} compared to those of  M3
          {\it (middle panel)} and M13 {\it (right panel)}. CMDs of M3 and M13 (from \citet{Chen+2021})
          have been shifted, respectively, by $\Delta m_{\rm F275W}=-0.35$ and  $\Delta m_{\rm F275W}=+0.20 $ in magnitude and by $\Delta (m_{\rm F275W}-m_{\rm F336W})=+0.13$ and  $\Delta (m_{\rm F275W}-m_{\rm F336W})=+0.07 $ in colour.}
	\label{fig:2}
\end{figure}

 The overall morphology of the CMD appears to be very similar to that observed in M3. This can be easily appreciated in Figure \ref{fig:2}, where the CMD of M3 (from \citealt{Chen+2021}) has been shifted to match the color and magnitude location of the evolutionary sequences of M5. Indeed, apart from the slightly different metal content ([Fe/H]$\sim -1.5$ and [Fe/H]$\sim -1.3$ for M3  and M5, respectively;  see \citealt{Harris+1996}) that introduces some second-order differences, the evolutionary sequences in the two clusters (M3 and M5) are impressively similar. In particular, the color extension of the HB is fully comparable, with the extremely blue part, at $(m_{\rm F275W}-m_{\rm F336W})<-0.1$, being essentially not populated in both cases.
For the sake of comparison, in the same figure we also included the CMD of M13 (from \citealt{Chen+2021}, shifted to match the color and magnitude location of the evolutionary sequences of M5), which instead shows a clearly different HB morphology with  a pronounced blue tail population. The impressive similarity of the HB morphologies in M5 and M3 suggests an analogous post-HB evolutionary path for the stars in these two clusters, surely different from that expected in M13. Thus, on the basis of the HB morphology  a population of canonical WDs (similar to that found in M3) and no presence of slowly cooling WDs (as those observed in M13) are expected in M5. 
  
\subsection{Artificial star tests}
\begin{figure}
\centering \includegraphics[width = 15cm]{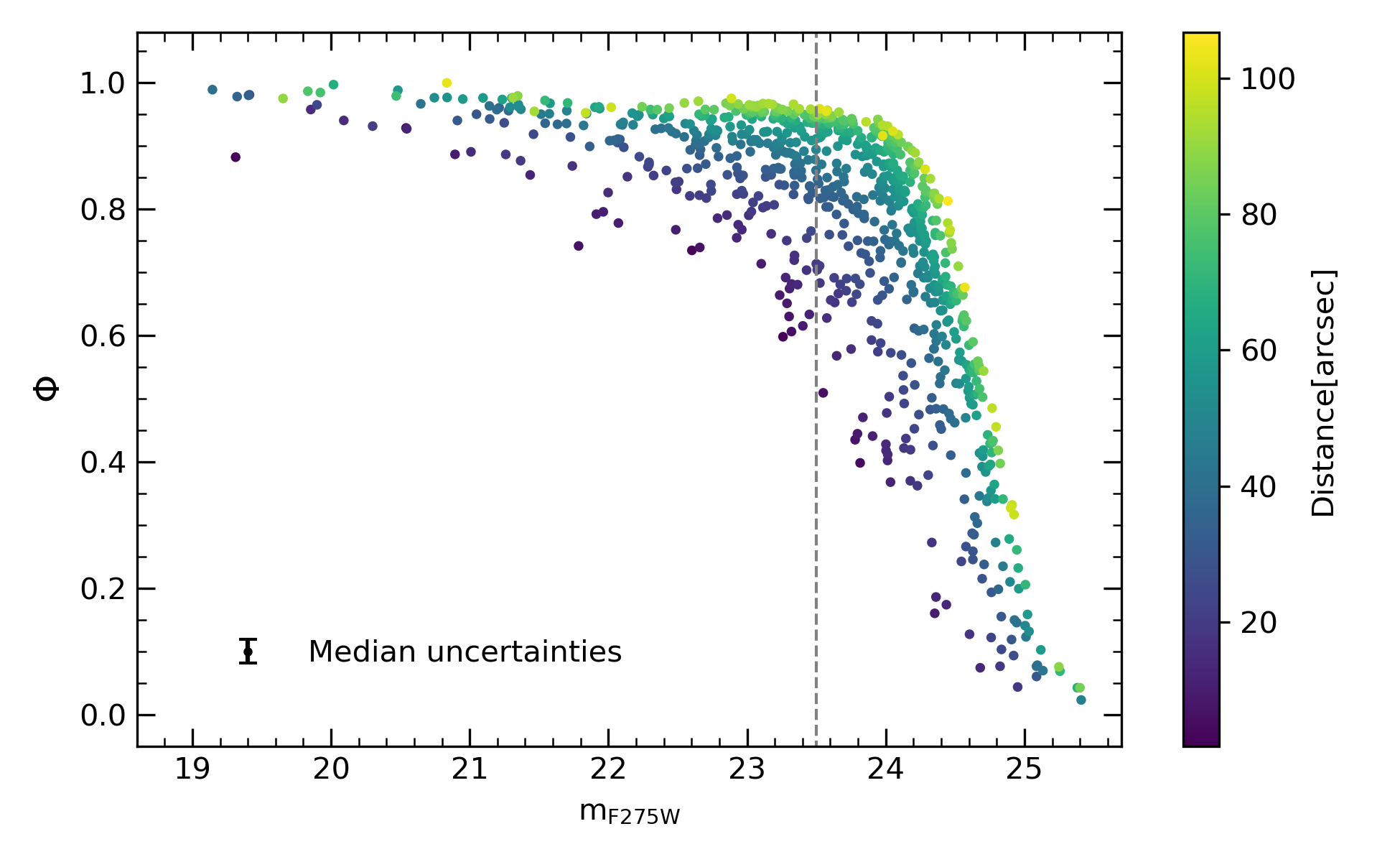}
	\caption{Completeness parameter $\Phi$ of the selected WDs as
          a function of the $m_{\rm F275W}$ magnitude, with the radial
          distance from the cluster center color-coded as indicated in
          the colorbar. The median uncertainty on $\Phi$ is marked in
          the panel. All the WDs located on the left of the vertical
          dashed line have completeness $\Phi>0.5$. }
	\label{fig:compl}
\end{figure}
A meaningful quantitative study of the properties
of the LF of faint objects like WDs requires, as a mandatory step, the empirical
determination of the level of photometric completeness of the WD
cooling sequence at different levels of magnitudes and distances from
the cluster center. This procedure should provide a first-order
estimate of the amount of sources lost during the data reduction
process because of their intrinsic faintness and the level of crowding
of the surrounding environment.  The evaluation of the photometric
completeness was performed by means of extensive artificial star
experiments. The standard recipe to perform these tests is described
in detail in \citet[][see also \citealt{Dalessandro+2015, Cadelano+2020b, Cadelano+2022b, Chen+2021}]{Bellazzini+2002}, thus here we just
summarize the main steps of the procedure adopted for M5. As first
step, we created a list of artificial stars with an input F275W
magnitude sampling the observed extension of the WD cooling
sequence. Figure~\ref{fig:1} shows that the observed WD cooling sequence
in M5 extends from $m_{\rm F275W}\approx 19$ to $m_{\rm F275W}\approx
25.5$. For each of these stars a corresponding magnitude in the F336W
filter was assigned according to the mean ridge line of the cooling
sequence in the $(m_{\rm F275W}, m_{\rm F275W}-m_{\rm F336W})$ CMD. A
large number of artificial stars were generated and then added to each
real image with DAOPHOT/ADDSTAR software. All artificial stars were
placed onto the images following a regular grid of $23\times23$ pixels
(corresponding to about 15 times the FWHM of the stellar
sources). Note that to avoid an artificial increase of the crowding
conditions, only one artificial star was arranged in each cell. The
entire photometric analysis was then repeated following exactly the
same procedure described in Sect.~\ref{sec:data}.  The procedure was
then iterated several times to ensure a sufficiently large sample and,
at the end, more than 160,000 artificial stars were simulated in the
entire field of view that covers a region within approximately
$90\arcsec$ from the cluster center.
 
To quantify the level of photometric completeness, the completeness
parameter $\Phi$ was determined as the ratio between the number
$N_r$ of artificial stars recovered by the photometric analysis, and the
number of stars that were actually simulated (number of input stars,
$N_i$). Of course the value of $\Phi$ is expected to be strongly
dependent on both crowding (hence, the distance from the cluster
center) and luminosity: it commonly decreases in the innermost regions
of the clusters (due to the large stellar density) and at faint
magnitudes.  To properly trace both these effects, we divided the
sample of simulated stars in radial bins at different distances from
the cluster center (with steps of $5\arcsec$) and in F275W magnitudes
bins (in steps of 0.5 mag) and, for each cell of this grid, we counted
the number of input and recovered stars, calculating the corresponding
value of $\Phi$.  The computed value of $\Phi$ was then assigned to
each WD according to its radial position and magnitude. The size of
the adopted magnitude and radial distance steps was set to guarantee a
reasonable statistics, while maintaining a sufficiently high spatial
resolution and sensitivity of the $\Phi$ parameter to the stellar
luminosity.  The uncertainties on the completeness value
($\sigma_\Phi$) were computed by propagating the Poisson errors,
and typically are on the order of 0.05.  The radial distances have
been computed with respect to the center of gravity quoted in
\citet[][see also \citealt{Cadelano+2017b, Miocchi+2013, Lanzoni+2010, Lanzoni+2019}
  for the details on the method adopted to estimate GC
  centers]{Lanzoni+2007}, which is located at right ascension
$\alpha_{\rm J2000} = 15^{\rm h}\, 18^{\rm m}\, 33\fs53$, and
declination $\delta_{J2000} = +02\arcdeg\, 04\arcmin\, 57\farcs06$,
with a $1\sigma$ uncertainty of $0\farcs 5$ in both the coordinates.
The construction of such a completeness grid allowed us to assign the
appropriate $\Phi$ value to each observed WD, with given F275W and
F336W magnitudes, and located at any distance from the cluster
center. The behavior of the completeness parameter as a function of
the magnitude is shown in Figure \ref{fig:compl} for all the detected
WDs.  According to previous works, we limited the analysis of the WD
LF to the brightest portion of the cooling sequence ($m_{\rm F275W}<
23.5$), where the completeness level is larger than 50\% at any
cluster-centric distance.


\section{Analysis}
\label{sec:analysis}

\begin{figure*}
	\centering
	\includegraphics[width=10cm]{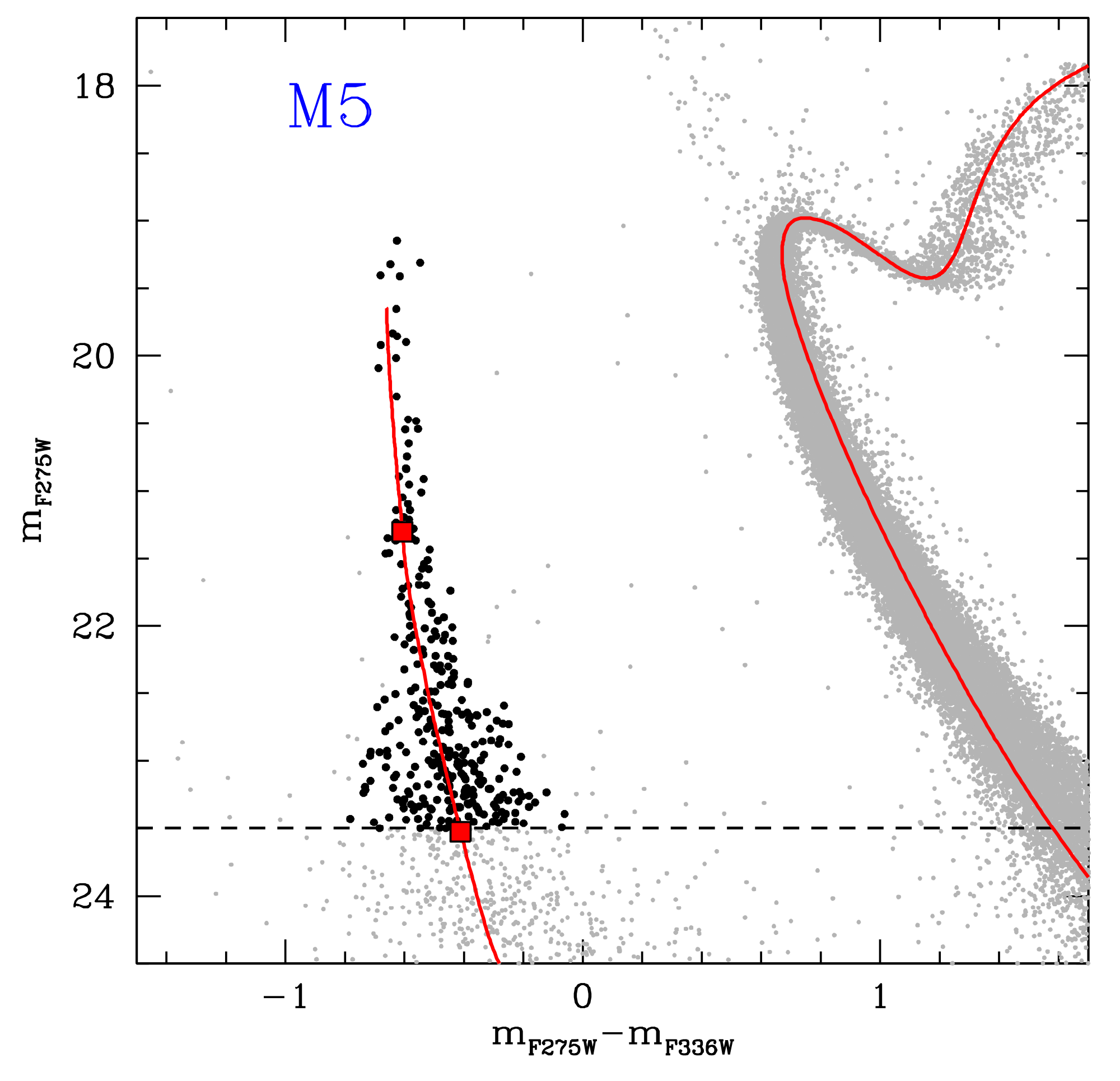}
	\caption{CMD of M5 zoomed in the WD cooling sequence
          region. The analyzed WDs are marked with black dots, and the
          cut-off magnitude is marked with the horizontal dashed line
          located at $m_{\rm F275W} = 23.5$. The red line is the
          cooling track of a $0.54 M_\odot$ WD, along which two
          reference cooling ages (10 and 100 Myr) are marked
          with red squares. The 12.5 Gyr old isochrone (with $Z=0.002$
          and helium mass fraction $Y=0.248$) from the BaSTI models
          (\citealp{Pietrinferni+2006}) well reproducing
          the cluster MS-TO. 
          }
	\label{fig:zoom}
\end{figure*}


\subsection{Sample selection and WD LF}
\label{sec:LF}
Figure \ref{fig:zoom} shows a portion of the CMD of M5 zoomed in the
WD region.
Overplotted to the data is the carbon-oxygen WD cooling track
\citep{Salaris2010} of a $0.54 M_\odot$ with hydrogen atmosphere, which 
nicely matches the observed cooling sequence even in
its brightest portion. For the sake of comparison we also overplotted
a 12.5 Gyr, $\alpha$-enhanced isochrone with metal abundance $Z=0.002$
and helium mass fraction $Y=0.248$ taken from the BaSTI models
(\citealp{Pietrinferni+2006}; see also \citealp{Hidalgo+2018,
  Pietrinferni+2021}), which well reproduces the MS-TO region of the
cluster.


For a proper study of the LF, we selected a sample of \emph{bona fide}
WDs by following the same procedure already adopted in previous papers
(see \citealt{Chen+2021, Chen+2022}): (1) we considered all the
objects located within 3$\sigma$ from the mean ridge line of the WD
cooling sequence, with $\sigma$ being the photometric error at the
corresponding magnitude level, and the mean ridge line essentially
corresponding to the $0.54 M_\odot$ WD track; (2) as discussed above,
we conservatively retained only WDs with completeness level above
$50\%$, thus limiting the sample to WDs brighter than $m_{\rm F275W}=23.5$ mag.
This magnitude cut corresponds to a cooling time of $\sim 100$ Myr,
which is comparable to the threshold adopted in the case of
M3 and M13 by \citealt{Chen+2022}.

\begin{figure*}
 	\centering
 	\includegraphics[width=14cm]{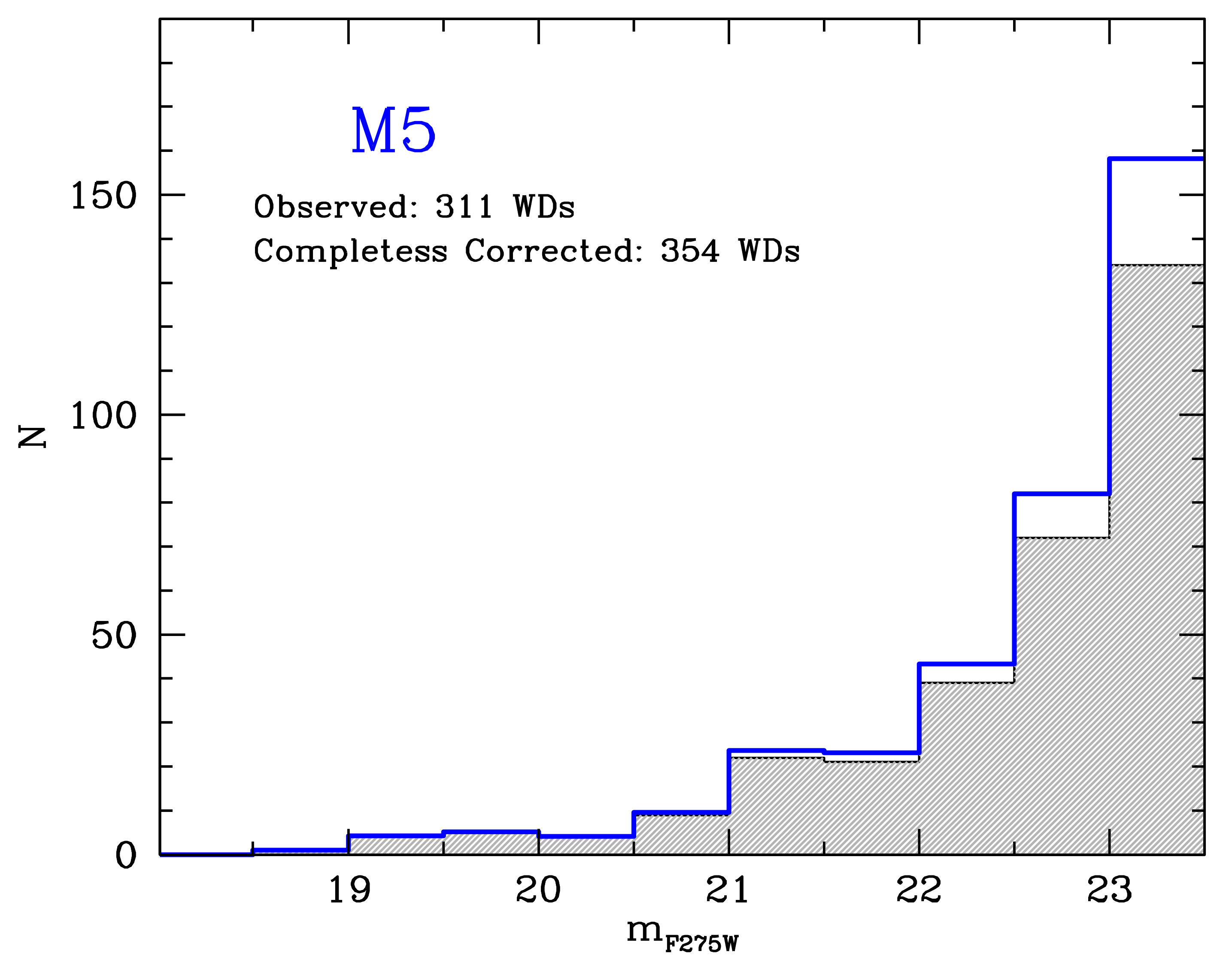}
 	\caption{Observed (grey shaded) and completeness-corrected
          (blue) LF of the selected WD sample in M5.}
 	\label{fig:LF}
 \end{figure*}

Following the adopted selection criteria we obtained a sample of 311 WDs
 in M5.  Their LF, computed in bins of 0.5 magnitudes,
 is shown in Figure \ref{fig:LF} for both the observed and the 
completeness-corrected samples (grey shaded and blue histograms,
respectively). It is worth noticing that the conservative criteria
adopted for the sample selection strongly limit the impact of
incompleteness: the global correction for completeness is 
smaller than $14\%$ (43 stars in total), with the
completeness-corrected population of WDs amounting to 354 stars.

\begin{figure*}
	\centering \includegraphics[width=14cm]{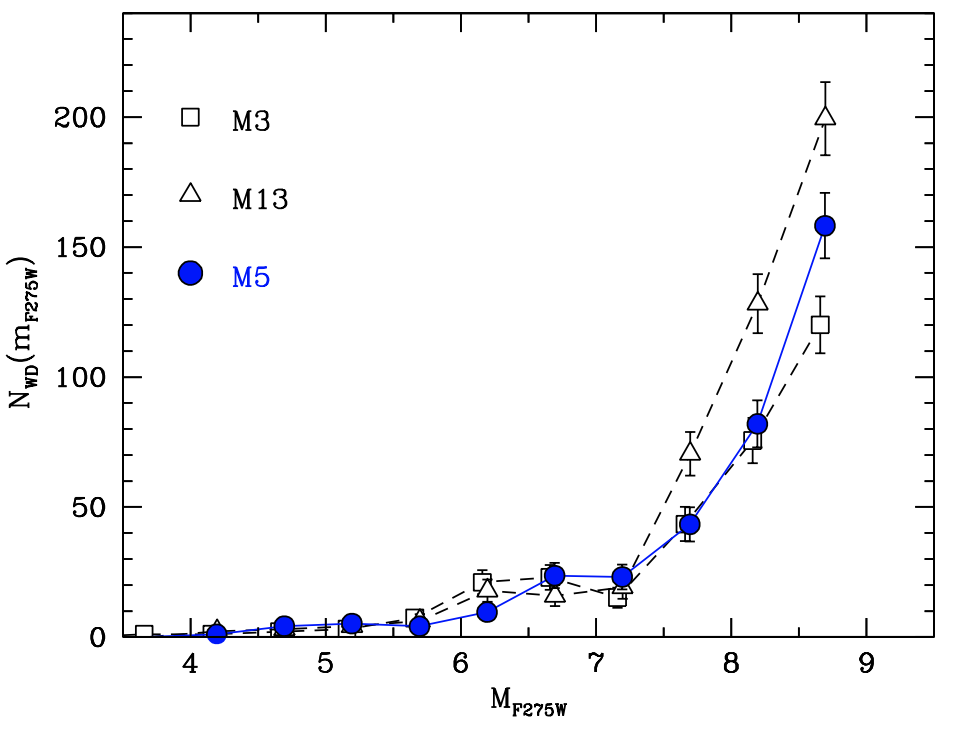}
	\caption{Completeness-corrected WD LF of M5 (blue circles),
          compared to that of M3 (empty squares) and M13 (empty
          triangles). The apparent $m_{\rm 275W}$ magnitude has been
          transformed in absolute magnitude by adopting distance
          modulus and reddening of each cluster (from
          \citealt{Ferraro+1999}; see also Table \ref{tab:para}). }
	\label{fig:cfr}
\end{figure*}


\subsection{Comparing WD LFs}
\label{sec:cfr}
We can now compare the completeness-corrected WD LF of M5 with those
previously obtained for the two twin clusters M3 and M13
\citep[see][]{Chen+2022}.  Intriguingly, both the limiting magnitude
and the total number of WDs selected in M5 are very similar to those
of M3. Indeed, this already suggests that the LFs observed in these
clusters can be compared directly, with no need for normalization
factors. In addition, by adopting the integrated apparent luminosity
quoted by \citet{Harris+1996}, and the distance modulus and reddening
values listed in \citet[][see also Table
  \ref{tab:para}]{Ferraro+1999}, we found that the integrated absolute
magnitudes of the three clusters are quite similar ($\sim -8.8$; see
Table \ref{tab:para}), denoting comparable total masses and
total numbers of stars.  According to this evidence, we then compared
the three WD LFs in terms of absolute star counts (with no
normalization) as a function of the absolute F275W magnitude ($M_{\rm
  F275W}$). Figure \ref{fig:cfr} shows that the LFs of M3
and M5 appear to be very similiar, and they are both clearly different
from the LF of M13.

\begin{figure*}
	\centering
	\includegraphics[width=14cm]{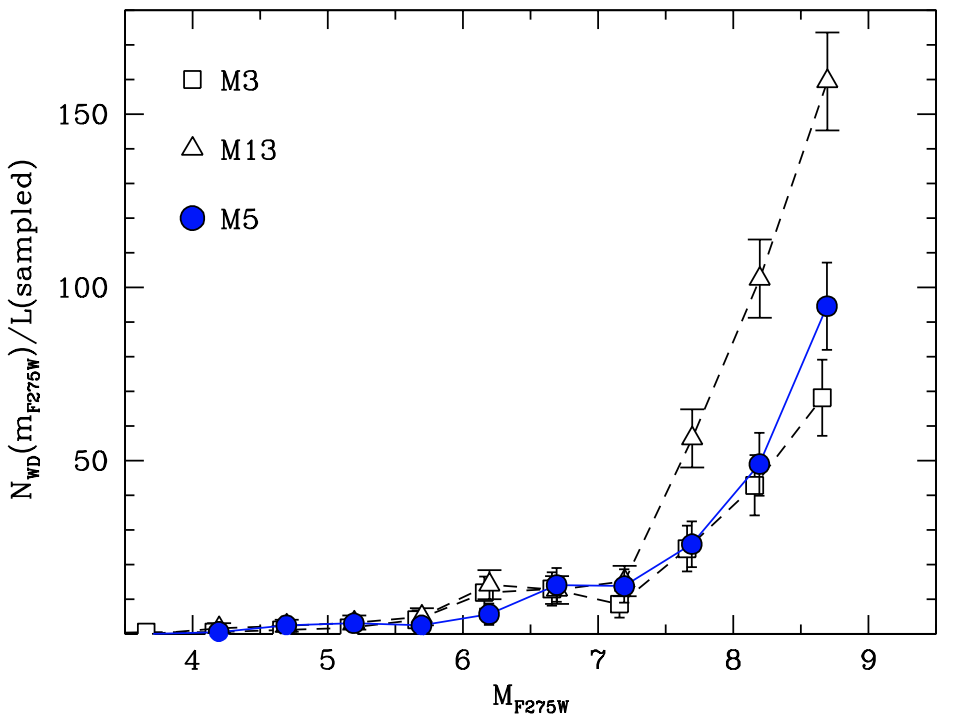}
	\caption{Completeness-corrected WD LFs of M5 (blue circles),
          M13 (empty triangles), M3 (empty squares) normalized to the
          sampled luminosity in units of $10^5 L_\odot$, as a function
          of the absolute $F275W$ magnitude. }
	\label{fig:cfr_norm}
\end{figure*}

To double check this result and for a more rigorous comparison (taking
into proper account the slightly different intrinsic richness of each
cluster), we then determined the total luminosity sampled by the
adopted observations.  The HST-WFC3 pointings sampled the innermost
$\sim 90\arcsec$ from the center of each cluster.  However, due to the
different intrinsic structures in terms of core radius and
concentration (\citealt{Miocchi+2013}; see Table \ref{tab:para}), this
may correspond to different fractions of the total luminosity. Thus,
we estimated the amount of total luminosity sampled at
$r<90\arcsec$ in each system by integrating the corresponding best-fit
King model \citet[from][]{Miocchi+2013}. The results are listed in the
last column of Table \ref{tab:para} and show that the differences are
admittely small (on the order of $10-15\%$). However, they go in the
direction of making the WD LF of M5 more similar to that of M3 and
increase the difference with the LF of M13, as clearly shown by 
Fig.\ref{fig:cfr_norm}, which displays the WD LFs normalized to the
sampled luminosities in units of $10^5 L_\odot$.
The number of WDs per $10^5 L_\odot$ sampled luminosity  with a cooling time $<100 $Myr
turns out to be approximately a half ($47-56\%$) of that counted in M13. 

\begin{table}
\centering
\caption{Photometric and structural parameters of M5, M13 and M3}
\begin{tabular}{cccccccc}
\hline
\hline
\ Cluster & $(m-M)_0$    &$E(B-V)$& $V_t$ &  $r_c$        & $c$  & $M_V$  & $L_{\rm sampled}$        \\
\hline
\ M5      & 14.37	 & 0.03   & 5.65  & $28\arcsec$	 & 1.66  & -8.81 & $1.67\times 10^5 L_\odot$\\
\ M3      & 15.03	 & 0.01	  & 6.19  & $22.7\arcsec$& 1.85  & -8.87 & $1.78\times 10^5 L_\odot$\\
\ M13     & 14.43	 & 0.02   & 5.78  & $49.5\arcsec$& 1.32	 & -8.71 & $1.25\times 10^5 L_\odot$\\
\hline
\end{tabular}
\tablecomments{$V_t$ is the total integrated $V-$band magnitude
  \citep[from][]{Harris+1996}; $(m-M)_0$ and $E(B-V)$ are,
  respectively, the distance modulus and reddening
  \citep[from][]{Ferraro+1999}; $M_V$ is the derived absolute
  magnitude; $L_{\rm sampled}$ is the luminosity sampled by the
  observations, obtained from the integration of the best fit King
  model to the star density profile of each system
  \citep[from][]{Miocchi+2013}.}
\label{tab:para}
\end{table}

\begin{figure*}
	\centering
	\includegraphics[width=14cm]{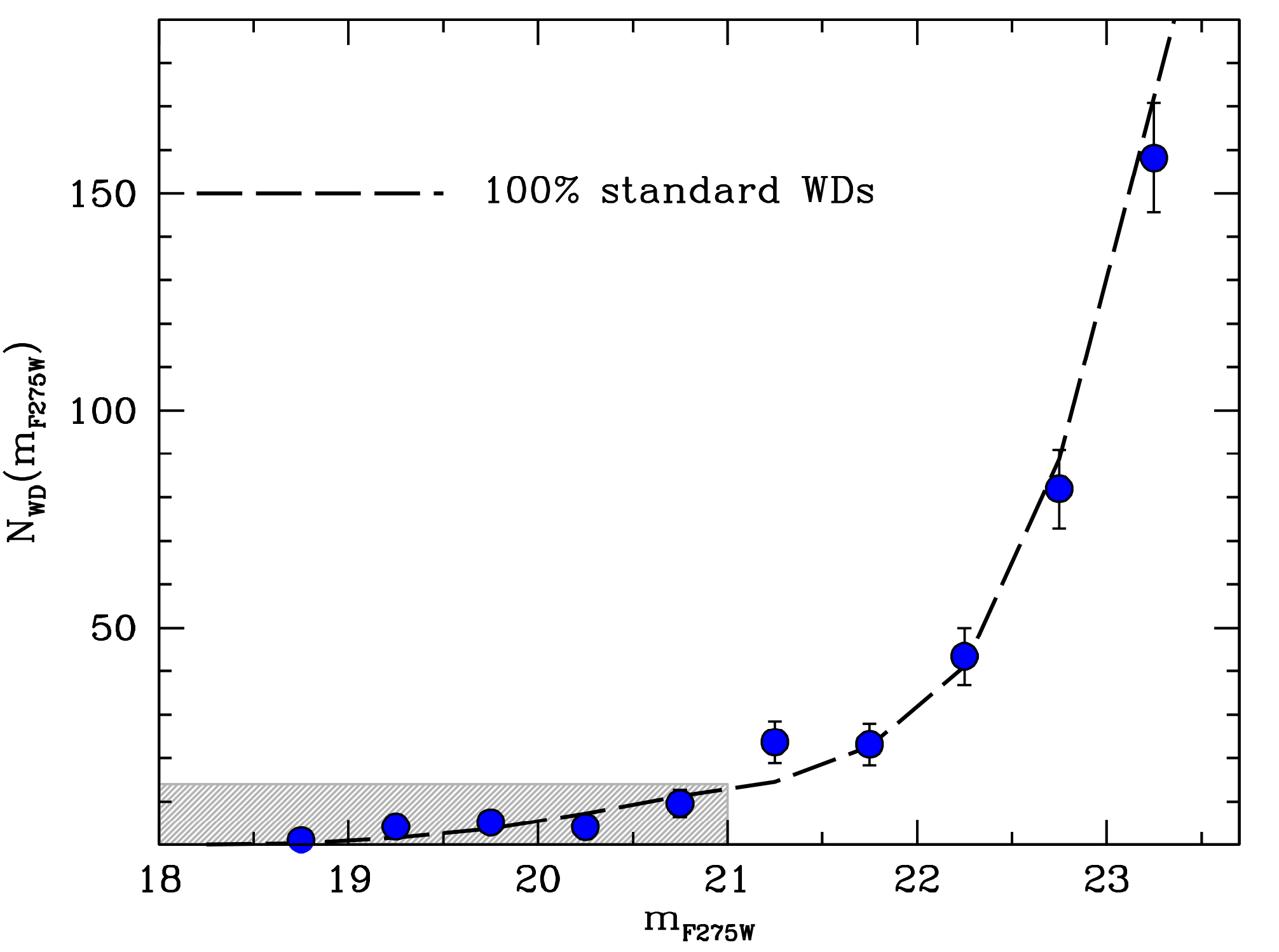}
	\caption{Completeness-corrected WD LF of M5 (blue circles),
          compared to the theoretical LF expected for a population of
          canonical (fast cooling) WDs (dashed line). The two LFs are
          normalized to the total number of WDs detected in the
          brightest portion of the cooling sequence ($m_{\rm
            F275W}<21$), indicated by the gray shaded region. }
	\label{fig:teo}
\end{figure*}


\section{Discussion and Conclusions}
\label{sec:con}
The scenario presented in \citet[][see also
  \citealt{Chen+2022}]{Chen+2021} suggests the existence of a special
class of \lq{slowly cooling WDs\rq}, whose cooling rate is slowed down by
some stable hydrogen burning in a very thin, but massive enough (of
the order of $10^{-4} M_\odot$; \citealt{Renedo+2010}) residual
hydrogen-rich outer layer.  The observational
signature of these objects is an excess of WDs in the brightest
portion of the LF, with respect to what observed in clusters where
they are not present, and to predictions from models of
\lq{canonical\rq} (fast cooling, with no hydrogen burning) WDs \citep[e.g.][]{Salaris2010}.
Such an excess was indeed detected in M13 and NGC 6752, but not in M3.

A convincing picture of the physical origin of these objects must take
into account that the mass of any residual hydrogen in proto-WDs is
regulated by the occurrence (or lack of) of the third dredge-up during the
AGB evolutionary phase \citep{Althaus+2015}.  In fact, during the
third dredge-up, convection carries carbon up to the stellar surface,
while hydrogen is mixed into inner regions where it is burned.
Thus, the occurrence of the AGB third dredge-up is expected to
generate proto-WDs with very small (in terms of mass) hydrogen envelopes, where
thermonuclear burning cannot take place: as a consequence, these stars
end up as canonical WDs.  Conversely, stars able to
skip the third dredge-up reach the WD stage with a residual hydrogen
envelope thick enough (with masses at least a few times $10^{-4} M_\odot$) to allow stable
thermonuclear burning, and this process provides the WD with an
extra-energy production that delays its cooling: slowly
cooling WDs are thus generated. \citet{Chen+2021} convincingly
demonstrated that the occurrence of this phenomenon is linked to the
cluster HB morphology: significant populations of slowly cooling WDs
are expected in GCs with well populated and extended blue HB
morphologies.  In fact, the presence of an extended HB blue tail
indicates the existence of a significant fraction of stars with
envelope mass so small that the subsequent AGB thermally pulsing phase
(hence, the third dredge-up) cannot occur.
Thus, in M13 and NGC 6752 (both having blue tail HBs) $\sim 70\%$ of
the HB stars are expected to completely or partially skip the AGB, 
guaranteeing the survival of a significant residual hydrogen
envelope and the consequent generation of slowly cooling WDs.
Conversely, in M3 (where the HB morphology does not display an extended blue
tail) essentially all the HB stars are expected to evolve along the
AGB, experience thermal pulses and third dredge-up, and thus produce canonical WDs.
 
From the analysis presented in this paper, the WD LF observed in M5
appears to be similar to that obtained by \citet{Chen+2021} in
M3. According to the scenario summarized above, the WD cooling
sequence of M5 is thus expected to be essentially populated by
canonical WDs, whose progenitors all experienced the third dredge-up
during the AGB phase. This is fully consistent with the expected mass
distribution along the HB in the two clusters, which appears
morphologically similar, with no extended blue tail (see Figure
\ref{fig:2} and \citealt{Dalessandro+2013a} for the HB mass function in
M3).

To further corroborate this conclusion, we have compared the observed
HB star distribution in both the UV and F606W-(F606W-F814W) optical
CMDs of M5, with theoretical models from the BaSTI database.
Specifically we used the $0.56 M_\odot$,
$\alpha$-enhanced HB tracks from \citet{Pietrinferni+2006} at the
cluster metallicity ([Fe/H]$=-1.3$) to identify the position of stars
with this mass along the HB of M5. The adopted value marks
approximately the lower limit of the HB masses that eventually go on
experiencing the third dredge-up.  To take into account the presence
of a range of He abundances up to $\Delta Y \sim 0.04$ (see Milone et
al. 2018), we employed tracks for both \lq{normal\rq} ($Y=0.248$) and
enhanced ($Y=0.30$) helium mass fractions.  We found that only 10\%
(at most) of the HB stars have mass smaller than $0.56 M_\odot$, thus
confirming that basically all cluster WDs are expected to be
canonically (fast) cooling objects.
 
Finally, we quantitatively tested this prediction by comparing the
completeness-corrected WD LF of M5 (blue circles in Figure
\ref{fig:teo}) with a theoretical WD LF computed in the case of 100\%
canonical WDs (dashed line in the figure).  The theoretical LF is 
normalized to have the same total number of objects as
in the brightest portion of
the observed LF (for $m_{\rm F275W}<21$: grey shaded region in the
figure).  As expected, the WD LF of M5 turns out to be in excellent
agreement with the theoretical sequence, clearly demonstrating that
the WD population in this cluster with no blue HB tail is entirely
constituted by canonical objects.

The results presented in this work therefore provide further solid
support to the scenario traced in \citet{Chen+2021, Chen+2022} about
the origin of slowly cooling WDs and their link with the HB morphology
of the parent cluster.  The extension of this investigation to
clusters with extended HB blue tail in the extreme low metallicity
regime ([Fe/H]$=-2.2$), where the phenomenon is expected to reach its
maximum efficiency, is now urged to fully verify the theoretical
predictions and provide an empirical measure of the impact of these
results on the use of the WD cooling sequences as chronometers to
measure cosmic ages.

\acknowledgments 
This research is part of the project {\it Cosmic-Lab} at
the Physics and Astronomy Department of the University of Bologna
(http://www.cosmic-lab.eu/Cosmic-Lab/Home.html). The research has been
funded by project {\it Light-on-Dark}, granted by the Italian MIUR through
contract PRIN-2017K7REXT (PI: Ferraro). JXC acknowledges the support
from China Scholarship Council (CSC).

\vspace{5mm}

\facilities{HST(WFC3)}
\software{DAOPHOT\citep{Stetson+1987}, DAOPHOT/ALLFRAME\citep{Stetson+1994}}


\vskip12pt


\end{document}